\documentclass{elsart} 
\journal{Nuclear Physics B} 
\usepackage{graphicx} 
\usepackage{amsmath} 

\newcommand{\la}{\label} 
\newcommand{\be}{\begin{equation}} 
\newcommand{\ee}{\end{equation}} 
\newcommand{\bea}{\begin{eqnarray}} 
\newcommand{\eea}{\end{eqnarray}} 

\def\la{\label} 
\def\be{\begin{equation}} 
\def\beq{\begin{equation}} 
\def\eeq{\end{equation}} 
\def\ee{\end{equation}} 
\def\bea{\begin{eqnarray}} 
\def\eea{\end{eqnarray}} 
\def\p{\partial} 

\begin{document} 

\begin{frontmatter} 

\title{Semiclassical evolution of the spectral curve in the normal 
random matrix ensemble as Whitham hierarchy} 

\author[la]{R.~Teodorescu\thanksref{now}}, 
\ead{rteodore@uchicago.edu} 
\author[lb]{E.~Bettelheim\thanksref{now1}}, 
\ead{eldadb@phys.huji.ac.il} 
\author[lb]{O.~Agam}, 
\ead{agam@phys.huji.ac.il} 
\author[lc]{A.~Zabrodin}, 
\ead{zabrodin@itep.ru} and 
\author[ld]{P.~Wiegmann} 
\ead{wiegmann@uchicago.edu} 
\thanks[now]{Present address: Columbia University, 
Department of Physics, 538 W. 120th St., New York, NY 10027, USA.} 
\thanks[now1]{Present address: James Frank Institute, University of Chicago, 
5640 S. Ellis Ave. Chicago, IL 60637, USA.} 
\address[la]{James Frank Institute, University of Chicago, 5640 S. Ellis Ave. 
Chicago, IL 
60637, USA.} 
\address[lb]{Racah Institute of Physics, Hebrew University, Givat Ram, 
Jerusalem, Israel 
91904} 
\address[lc]{Institute of Biochemical Physics, Kosygina str. 4, 117334 Moscow, 
Russia, also at ITEP, Bol. Cheremushkinskaya str. 25, 
117259 Moscow, Russia} 
\address[ld]{James Frank  Institute, Enrico Fermi Institute, University of 
Chicago, 5640 S. Ellis Ave. Chicago, IL 60615, USA, and the Landau 
Institute, Moscow, Russia} 

\begin{abstract} 
We continue the analysis of the spectral curve of the 
normal random matrix ensemble, introduced in an earlier 
paper.  
Evolution of the full quantum curve is given in terms of 
compatibility equations of independent flows. The semiclassical 
limit of these flows is expressed through canonical differential 
forms of the spectral curve. We also prove that the semiclassical 
limit of the evolution equations is equivalent to Whitham hierarchy. 
\end{abstract} 


\end{frontmatter} 

\section{Introduction} 
       \label{sec:intro} 

Continuing the study of the normal random matrix ensemble, 
presented in the series of papers \cite{mwz,ABWZ02,mkwz,TBAZW04}, 
we explore the evolution of the spectral curve  with respect to all the 
independent flows. 

In the first part of the paper, we review the basic concepts of the theory, 
relevant to many different developments, from integrability of non-linear 
differential equations to supersymmetric Yang-Mills theory \cite 
{DV03,BEH,Kapaev,gravity,Its,KM03,Chekhov}. 

In the second part, we derive the 
evolution equations and prove that their semiclassical limit 
is the universal Whitham hierarchy associated with the complex curve. 
We also notice that the Whitham hierarchy is identical to the set of equations 
which describe Laplacian growth processes -- 
unstable dynamics of an interface 
between two immiscible phases. 

\subsection{Normal matrix ensemble} 

The following is a short review of normal random matrix theory. 
Normal matrices $M$ commute with their hermitian conjugate: 
$[M, M^{\dag}]=0$, so that both $M$ and $M^{\dag}$ 
can be diagonalized simultaneously, and have  complex eigenvalues. 
The statistical weight of the normal matrix ensemble is given through a 
general potential $W(M,M^{\dag})$ \cite{Zaboronsky}: 
\begin{equation} 
\label{ZN} 
e^{\frac{1}{\hbar}{\rm tr} \, W(M, M^{\dag})} d\mu (M). 
\end{equation} 
Here $\hbar$ is a parameter, and the measure of integration 
over normal matrices 
is induced by the flat metric 
on the space 
of all complex matrices. 
Using the standard procedure \cite{Mehta91}, angular degrees of freedom 
are integrated out, leading to the joint 
probability distribution 
of eigenvalues $z_1,\dots,z_N$, where $N$ is 
size of the matrix: 
\begin{equation} 
\label{mean} 
\frac{1}{N! \tau_N}|\Delta_N (z)|^2 \,\prod_{j=1}^N 
e^{\frac{1}{\hbar}W(z_j,\bar z_j)} d^2 z_j. 
\end{equation} 
Here 
$\Delta_N(z)=\det (z_{j}^{i-1})_{1\leq i,j\leq N}= 
\prod_{i>j}^{N}(z_i -z_j)$ 
is the Vandermonde determinant, and 
\begin{equation} \label{tau} 
\tau_N = \frac{1}{N!}\int 
|\Delta_N (z)|^2 \,\prod_{j=1}^{N}e^{\frac{1}{\hbar} 
W(z_j,\bar z_j)} d^2z_j 
\end{equation} 
is a normalization factor, the partition function 
of the matrix model.

We consider the case where the potential $W$ has the form 
\begin{equation} 
\label{potential} 
W=-|z|^2+V(z)+\overline{V(z)}, 
\end{equation} 
where $V(z)$ is a holomorphic function in a domain which 
includes the support of eigenvalues. 
We also assume that the field 
\be\la{a} 
A(z)=\p_z V(z) 
\ee 
(a ``vector potential'') is a globally defined meromorphic function. 

     In a proper large $N$ limit ($\hbar\to 0$, 
$N\hbar$ fixed), the eigenvalues of matrices 
occupy  a connected domain $D$ 
       in the complex plane, or, in general, several 
disconnected domains. We refer to the connected components 
$D_\alpha$ of the domain $D$ as {\it droplets}.

In the case of algebraic domains \cite{TBAZW04}, 
the eigenvalues are distributed with the density 
$\rho=-\frac{1}{4 \pi}\Delta W$, 
where $\Delta = 4\p_z \p_{\bar z}$ is the 2-D Laplace operator \cite{WZ03}. 
For the potential (\ref{potential}), the density is uniform. 

Boundary components of the droplets form a real section of 
a complex curve, defined in the following way: let us 
represent the boundary of the domain 
as a real  curve $F(x,\,y)=0$. 
If the vector potential $A(z)$ is a meromorphic 
function (we always assume that this is the case), the function 
$F$ can be chosen to be an irreducible polynomial. 
Then we rewrite it 
in holomorphic coordinates as 
\be\label{F} 
F\left (\frac{z+\bar z}{2},\frac{z-\bar z}{2i}\right )= 
f(z,\bar z) 
\ee 
and treat $z$ and $\bar z$ as independent complex coordinates 
$z$, $\tilde z$. 
The equation 
$f(z,\tilde z)=0$ defines a complex curve. 
This curve is a finite-sheet covering of the $z$-plane. 
The single-valued function $\tilde z(z)$ 
on the curve is a multivalued function on 
the $z$-plane. Making cuts, one can fix single-valued 
branches of this function. The boundary of the domain 
is a section of the curve by the plane where 
$\tilde z$ is complex conjugate of $z$.

Similarly to the hermitian matrix ensemble 
\cite{DV03,KM03,Chekhov}, 
the complex  curve of the normal matrix ensemble 
   is characterized by the potential $W$ (or by 
the ``vector potential" 
$A(z)$), and  by a set of $g+1$ 
integers $\nu_\alpha$ (not necessarily positive), where $g$ is 
genus of the curve. 
The integers are subject to the constraint 
$\sum_{\alpha=0}^g\nu_\alpha=N$. 
If they are all positive, then they are proportional to the areas 
of the droplets 
of uniformly distributed eigenvalues. In this case, every droplet contains 
$\nu_\alpha$ eigenvalues. 

As one varies the potential 
and the filling factors $\nu_\alpha$, 
the curve and the interface bounding the droplets evolve. 
Parameters of the potential 
(for example, poles and residues of the meromorphic function (\ref{a})) 
and filling factors are 
{\it deformation parameters } and {\it parameters of growth}. 
They are coordinates 
in the moduli space of the complex curves. 
An infinitesimal variation of the potential 
generates  correlation functions of the  ensemble  \cite{WZ03}.

\section{Spectral curve and wave functions}\la{1A} 
In this section we specify the potential to be 
of the form (\ref{potential}), and introduce the 
set of biorthogonal wave functions 
\be \la{chin} 
\psi_n (z)= 
e^{-\frac{|z|^2}{2\hbar}+\frac{1}{\hbar}V(z)}P_n (z),\quad\mbox{and} 
\quad\chi_n (z)= 
e^{\frac{1}{\hbar}V(z)}P_n (z), 
\ee 
where the holomorphic 
functions $\chi_n(z)$ 
are orthonormal in the complex plane 
with the weight 
$e^{-|z|^2/\hbar}$. 
Like traditional orthogonal polynomials, the bi\-ortho\-gonal polynomials 
$P_n$ (and the corresponding wave functions) 
obey a set of differential equations with 
respect to the argument $z$, and recurrence relations with respect to 
the degree $n$. Similar equations 
for  two-matrix models are discussed in  numerous papers 
(see, e.g., \cite{Aratyn}). 

In the basis (\ref{chin}), multiplication by $z$ 
is represented by the $L$-operator (the Lax operator): 
\begin{equation}\la{L1} 
L_{nm}\chi_m(z)=z\chi_n(z) 
\end{equation} 
(summation over repeated indices is implied). 
Obviously, $L$ is a lower triangular matrix with one 
adjacent upper diagonal, $L_{nm}=0$ as $m>n+1$. 
Similarly, the differentiation 
$\p_z$ is represented by an 
upper triangular matrix with one adjacent lower diagonal. 
Integrating by parts the matrix elements of the $\p_z$, one finds: 
\begin{equation}\la{M} 
(L^{\dag})_{nm}\chi_m = 
\hbar\p_z\chi_n, 
\end{equation} 
where $L^{\dag}$ 
is the hermitian conjugate operator. 

The matrix elements of $L^{\dag}$ are 
$$(L^{\dag})_{nm}=\bar L_{mn}=A(L_{nm})+ 
\int e^{\frac{1}{\hbar}W}\bar P_m(\bar z)\p_z P_n(z) d^2z, 
$$ 
where the last term is a lower triangular matrix.  The latter can be written 
       through negative powers of the Lax operator. 
Writing $\p_z\log P_n(z)=\frac{n}{z}+\sum_{k>1}v_k(n)z^{-k}$, 
one represents $L^{\dag}$ in the form 
\begin{equation}\la{M1} 
L^{\dag}=A(L)+(\hbar n) L^{-1}+\sum_{k>1}v^{(k)}L^{-k}, 
\end{equation} 
where $v^{(k)}$ and $(\hbar n)$ are 
diagonal matrices with elements $v_n^{(k)}$ 
and $(\hbar n)$. 
The coefficients $v_{n}^{(k)}$ are determined by the condition that 
lower triangular matrix elements of $A(L_{nm})$ are cancelled. 

In order to emphasize the structure of the operator $L$, we 
write it in the basis of the 
shift operator 
$\hat w$ such that $\hat w f_n =f_{n+1}\hat w$ for any 
sequence $f_n$. Acting on the wave function, we have: 
$$\hat w 
\chi_n=\chi_{n+1}.$$ 
In the $n$-representation, the operators $L$, $L^{\dag}$ 
acquire the form 
\begin{equation}\la{M11} 
L=r_n \hat w+\sum_{k\geq 0} u_{n}^{(k)} \hat w^{-k},\quad 
L^{\dag} = \hat w^{-1} r_n+ 
\sum_{k\geq 0} \hat w^{k} \bar u_{n}^{(k)}. 
\end{equation} 
Acting on $\chi_n$, we have the commutation 
relation (``the string equation") 
\begin{equation}\la{string} 
[L, \, L^{\dag}]=\hbar. 
\end{equation} 
This is the compatibility condition of Eqs. (\ref{L1}) and (\ref{M}). 

Equations (\ref{M11}) and (\ref{string}) 
completely determine the coefficients 
$v_n^{(k)}$, $r_n$ and $u_n^{(k)}$. The first one connects the coefficients 
to the parameters of the potential. 
The second equation is used to determine how the coefficients 
       $v_n^{(k)}$, $r_n$ and $u_n^{(k)}$ evolve with $n$. In particular, 
its diagonal part reads 
\be\la{Area} 
n\hbar=r_n^2-\sum_{k\geq 1}\sum_{p=1}^k|u_{n+p}^{(k)}|^2. 
\ee

\subsection{Finite dimensional reductions} 
If the vector potential $A(z)$ is a rational function, 
the coefficients $u_{n}^{(k)}$ are not all independent. 
The number of independent coefficients 
equals the number of independent parameters of the potential. 
For example, if the holomorphic part 
of the potential, $V(z)$, is a polynomial of degree $d$, 
the series 
(\ref{M11}) are truncated at $k= d-1$.

In this case,  the semi-infinite system of linear equations (\ref{M}) 
and the recurrence relations (\ref{L1}) 
can be cast in the form of a set of finite dimensional differential equations, 
whose 
coefficients are rational functions of  $z$, one 
system for every $n>0$. 
The system of differential equations generalizes the 
Cristoffel-Daurboux second order 
differential equation 
valid for orthogonal polynomials. This fact has been observed in 
recent papers \cite{BEHdual,Eynard03} 
for biorthogonal polynomials emerging in the 
hermitian two-matrix model 
with a polynomial potential. It is 
applicable to our case (holomorphic biorthogonal 
polynomials) as well.

The semi-infinite set $\{\chi_0,\chi_1,\dots\}$ is 
then decomposed as a ``bundle" of 
$d$-dimensional vectors 
$${\underline\chi}(n)= (\chi_n,\chi_{n+1},\dots, 
\chi_{n+d-1})^{{\rm t}}$$ 
(the index ${\rm t}$ 
means transposition, so 
${\underline\chi}$ is a column vector). 
The dimension of the vector is the number of poles 
of $A(z)$ plus one. 
Each vector obeys a closed $d$-dimensional linear 
differential equation 
\begin{equation} 
\la{M2} 
\hbar\p_z{\underline\chi}(n)={\mathcal L}_n (z){\underline\chi}(n), 
\end{equation} 
where the $d\times d$ matrix 
${\mathcal L}_n$ is a ``projection" of the operator $L^{\dag}$ 
onto the $n$-th $d$-dimensional space. Matrix elements of the 
${\mathcal L}_n$ are rational functions of 
$z$ having the same poles as $A(z)$ and also a pole at 
the point $\overline{A(\infty )}$. If $A(z)$ is a polynomial, 
all these poles accumulate to a multiple pole at infinity.

\subsection{Spectral curve}\la{Curve1}

The semiclassical asymptotics of solutions to 
Eq. (\ref{M2}), as $\hbar \to 0$, 
are found by solving the eigenvalue problem for 
the matrix ${\mathcal L}_n (z)$ \cite{Wasow}. 
More precisely, the basic object of the the asymptotic analysis is the 
spectral curve \cite{curve} of the matrix ${\mathcal L}_n$, which is 
defined, for every integer $n>0$, by the secular equation 
$\det ({\mathcal L}_n (z)-\tilde z) = 0$ 
(here $\tilde z$ means $\tilde z \cdot {\bf 1}$, where 
${\bf 1}$ is the unit $d\times d$ matrix). 
It is clear that the left hand side of the secular 
equation is a polynomial in $\tilde z$ of degree $d$. 
We define the spectral curve by an equivalent equation 
\begin{equation}\la{qc} 
f_n (z,\tilde z)=a(z)\det ({\mathcal L}_n (z)-\tilde z) = 0, 
\end{equation} 
where the factor $a(z)$ is added 
to make $f_n(z,\tilde z)$ a polynomial in $z$ as well. 
The factor $a(z)$ then has zeros at the points where 
poles of the matrix function ${\mathcal L}(z)$ are located. 
It does not depend on $n$.

The spectral curve (\ref{qc}) is characterized by the 
antiholomorphic involution. 
In the coordinates $z, \tilde z$, the involution reads 
$(z, \tilde z)\mapsto (\overline{\tilde z}, \bar z)$. 
This simply means that the secular equation 
$\det (\bar {\mathcal L}_n (\tilde z)-z) = 0$ 
for the matrix $\bar {\mathcal L}_n (\tilde z)\equiv 
\overline{{\mathcal L}_n (\overline{\tilde z})}$ 
defines the same curve. 
Therefore, the polynomial $f_n$ 
takes real values for $\tilde z =\bar z$: 
\be\la{anti1} 
f_n(z,\bar z)=\overline{f_n ( z, \bar z)}. 
\ee 
Points of the real section of the curve 
($\tilde z =\bar z$) are 
fixed points of the involution. 

\subsection{Schwarz function} 

The polynomial $f_n(z, \bar z)$ can be 
factorized in two ways: 
\be\la{h1}f_n(z,\bar z)=a(z)(\bar z-S_n^{(1)}(z)) 
\dots (\bar z-S_n^{(d)}(z)), 
\ee 
where $S_n^{(i)}(z)$ are eigenvalues of the matrix 
${\mathcal L}_n (z)$, or 
\be\la{ah1} 
f_n(z,\bar z)=\overline{a(z)}( z-\bar S_n^{(1)}(\bar z))\dots ( z-\bar 
S_n^{(d)}(\bar z)), 
\ee 
where $\bar S_n^{(i)}(\bar z)$ are eigenvalues of the matrix 
$\bar {\mathcal L}_n (\bar z)$. 
One may understand them as different branches of a 
multivalued function $S(z)$ (respectively, $\bar S(z)$) 
on the plane (here we do not indicate the dependence on $n$, 
for simplicity of the notation). 
It then follows that 
$S(z)$ and $\bar S(z)$ 
are mutually inverse functions: 
\begin{equation}\la{anti11} 
\bar S(S(z))=z. 
\end{equation}

An algebraic function  with this property 
is called {\it the Schwarz function}. 
The equation $f(z, S(z))=0$ defines a complex curve 
with an antiholomorphic involution. An upper bound 
for genus of this curve is $g=(d-1)^2$, where $d$ is the 
number of branches of the Schwarz function. 
The real section of this curve is the set of all fixed 
points of the involution. It consists of a number 
of contours on the plane (and possibly a number of 
isolated points, if the curve is not smooth). 
The structure of this set is known to be 
  complicated. Depending on the 
coefficients of the polynomial, the number of disconnected 
contours in the real section may vary from $0$ to $g+1$. 
If the contours divide the complex curve into two 
disconnected ``halves", or sides (related by the involution), then 
the curve can be realized as 
the {\it Schottky double} \cite{Alhfors50,C,SS} of one of 
these sides. Each side is a Riemann surface with a boundary. 

The Schwarz function on the physical sheet 
is a particular root, say $S^{(1)}_n(z)$,   of the 
polynomial $f_n(z, \tilde z)$ (see (\ref{h1})). 
It follows from (\ref{M1}) that 
this root  is 
selected by the requirement that 
it has the same poles and residues as the potential $A$.

The formal\footnote{This formal expression ignores 
the Stokes phenomenon.} 
semiclassical asymptote of equation (\ref{M2}), 
in the leading order in $\hbar$, is 
\be\la{semi1} 
\chi_n\sim e^{\frac{1}{\hbar}\int^z d\Omega^{(1)}_n}. 
\ee 
Here 
$$d\Omega_n^{(1)}=S_n^{(1)}dz.$$ 
The differential $d\Omega^{(1)}$ is a physical branch of the generating 
differential on the curve (see below). 

The semiclassical asymptotics was 
discussed in more details in \cite{TBAZW04}. To next order in 
$\hbar$, it reads 
\be \la{75} 
\psi (z)\sqrt{dz}\sim \sqrt{W^{(\infty,\bar\infty )}(z)}\, 
e^{-\frac{1}{\hbar}\big(\frac{|z|^2}{2}-\int^z_{\xi_0} S(z) dz\big)}. 
\ee

\subsection{The generating differential}\la{Diff} 

The meromorphic differential 
\be\la{Omega} 
d\Omega=S(z)dz 
\ee 
plays an important role. 
It is called generating 
differential \cite{mkwz,kkmwz}. 
On the physical sheet 
it has the same poles and residues 
as the differential $Adz$: 
\be\la{1234} 
d\Omega=Adz+S_-(z)dz. 
\ee 

In the following, we use the following properties of the 
generating differential. 
\begin{itemize} 
\item [(i)] 
The periods over $\bf a$-cycles 
(boundaries of the droplets) are 
purely imaginary, and are integer multiples 
of $2\pi i$. They compute   areas 
of the droplets: 
$$ 
\nu_{\alpha}=\frac{1}{2\pi i \hbar}\oint_{{\bf a}_{\alpha}} 
d\Omega. 
$$ 
The filling factors of physical droplets 
(belonging to the physical sheet) 
are positive. 
\item[(ii)] 
The real part of the integral of the differential 
$(\bar z-S(z))dz$ from some fixed 
point $\xi_0$ to a point on the boundary 
of  a droplet (a point on a $\bf a$-cycle) 
     has the same value for all points of the boundary: 
\be \nonumber 
\phi_\alpha=-|z|^2+2 {\mathcal R}e\int_{\xi_0}^z d\Omega= 
\mbox{const,\quad\quad for all}\quad z\in {\bf a}_\alpha. 
\ee 
This quantity does not depend on $z$, 
but does depend on $\xi_0$ unless $\xi_0$ is on the boundary. 
However, the difference $\phi_{\alpha}-\phi_{\beta}$ 
depends on the ${\bf a}$-cycles only. 
It is equal to a $\bf b$-period of the differential 
$d\Omega$: 
     \be\la{32} 
\phi_\alpha-\phi_\beta=\oint_{{\bf b}_{\alpha\beta}}d\Omega, 
\ee 
where ${\bf b}_{\alpha\beta}$ is a 
cycle connecting ${\bf a}_{\alpha}$ and 
${\bf a}_{\beta}$ cycles. 
\end{itemize} 
For proofs and more details, see Ref. \cite{mkwz}. 

Periods over $\bf b$-cycles $\phi_\alpha-\phi_0$  play a role of 
chemical potentials for the filling factors. 
Here $0$ denotes a chosen reference droplet. 
One can use chemical potentials to 
characterize evolution of the curve instead 
of filling factors.

\section{Evolution of the spectral curve} 

\subsection{Evolution of the quantum curve} 

Evolution of wave functions of the matrix ensembles  with respect to 
a change of the potential is the subject of vast 
literature.  Most of it deals with orthogonal polynomials of  the 
hermitian  ensemble (see, e.g., \cite{Morozov}). 
Our case is similar to the case of 
biorthogonal polynomials 
appearing in two-matrix ensembles \cite{Aratyn}. 
We represent the evolution equations through variational 
derivatives $\delta /\delta A(z)$ 
with respect to the holomorphic part of 
the potential. This does not imply any particular 
parametrization of the potential. 
The  standard result reads (cf. \cite{kkmwz}): 
\begin{equation}\la{L2} 
\hbar \frac{\delta}{\delta A(\zeta)}\psi_n (z) 
=H_{nm}(\zeta)\psi_m(z), 
\end{equation} 
where 
\begin{equation} 
\la{L} 
H(\zeta )= 
-\left [\log \left (\zeta-L\right ) \right ]_{+} 
-\frac{1}{2} 
\left [\log \left (\zeta-L\right ) \right ]_0-\frac{1}{2} 
\log \zeta, 
\end{equation} 
and the notation $[\ldots ]_{+, 0}$ means 
the upper triangular or diagonal part of the matrix. This formula reflects 
the triangular structure of the $L$-operator encoded by Eq.(\ref{M1}) 
\begin{equation} \nonumber 
L^{\dag}=A(L)+(\hbar n) L^{-1}+\sum_{k>1}v^{(k)}L^{-k}. 
\end{equation} 
The compatibility equations 
\begin{equation}\la{ZS} 
\hbar \frac{\delta}{\delta A(z)} 
H(z')- \hbar \frac{\delta}{\delta A(z')} 
H(z)=[H(z),\,H(z')] 
\end{equation} 
determine evolution of $L$ 
(i.e., the coefficients $u_n^{(k)}$ in (\ref{M11})). 

Eqs. (\ref{L2},\ref{ZS}) describe an evolution with respect to 
deformation parameters,  while Eq. (\ref{string}) 
describe the growth (increasing $N$ while keeping all harmonic 
moments fixed). 

Equations (\ref{L1},\ref{L2},\ref{L},\ref{ZS}) constitute the 
Lax-Sato form of the Toda lattice integrable hierarchy \cite{Toda}. 
Biorthogonal polynomials form a very  particular solution of the 
hierarchy. 
If 
$A(z)$ is a rational function, the matrices 
$H(z)$ acquire a special structure, which is determined 
by poles and residues of $A(z)$. This case corresponds to 
finite-dimensional reductions of the Toda lattice hierarchy.

\subsection{Evolution of the semiclassical curve. Whitham hierarchy}\la{Ev} 
The variation of the Schwarz function, or the 
generating differential (\ref{Omega}), 
under a change of the deformation parameters  is a subject 
of deformation equations. 
The deformation equations are 
already built in the decomposition of the Schwarz 
function (\ref{1234}). 

In order to write them, 
we need holomorphic and meromorphic 
differentials canonically normalized 
with respect to the $\bf a$-cycles; they are defined in the following: 
\begin{itemize} 
\item [(i)] 
Holomorphic differentials 
(Abelian differentials of the first kind) $dW_\alpha(z)$, 
$\oint_{{\bf a}_\alpha} dW_{\beta}=\delta_{\alpha\beta}$; 
\item[(ii)] 
Meromorphic differentials 
(Abelian differentials of the third kind) $dW^{(\zeta,\bar\zeta)}(z)$ 
having poles at $z=\zeta$ and its mirror 
     on the back side with the residues $\pm 1$, normalized so that 
$\oint_{\bf a}dW^{(\zeta,\bar\zeta)}(z)dz=0$. 
The meromorphic differential  $dW^{(\infty,\bar\infty)}(z)$ 
plays a special role. 
\end{itemize}

The variation of $d\Omega =S dz$ with respect to $t=\hbar N$ 
gives the meromorphic differential 
$dW^{(\infty,\bar \infty )}(z)$: 
\be\la{34} 
\p_t S(z)dz=-dW^{(\infty,\bar \infty )}(z). 
\ee 
This follows from the fact that the singular part of the Schwarz 
function ($A(z)$ in (\ref{1234})) does not depend on $t$,  and  that 
$S_-(z)\to  t/z$ at infinity. 

Variations with respect to the filling factors 
$t^{(\alpha )}=\hbar \nu_\alpha$ at a fixed $t$ 
and a fixed potential affects only the $S(z)-t/z$ part of the Schwarz function. 
It gives some $g$- 
holomorphic differentials. We call them $dW_{\alpha}(z)$: 
\be\nonumber 
\p_{t^{(\alpha )}} S(z)dz=-dW_{\alpha}(z). 
\ee 
These differentials are canonically normalized: $\oint_{{\bf a}_\alpha} 
dW_{\beta}=\delta_{\alpha\beta}$, so they are unique. 

Finally, a variation of $d\Omega=Sdz$ with respect to 
$A(\zeta)$, at fixed filling factors, produces a 
meromorphic differential which has 
simple poles at $z=\zeta,\,\infty$ and their mirrors 
on the back side. The poles come from the variation of the first term of 
(\ref{1234}). Their residues are $\pm 1$ and the sum of the residues 
is zero on each side  of the double. Since all filling factors 
are fixed, this differential has zero ${\bf a}$-periods and 
we get 
\be\la{331} 
\frac{\delta}{\delta A(\zeta)}S(z)dz=-dW^{(\zeta,\bar\zeta )}(z)+ 
dW^{(\infty,\bar\infty )}(z),\quad\quad\mbox{on the front side}. 
\ee 
This equation simply reflects the fact 
that all singularities of the  Schwarz function are the singularities 
of  $A(z)$.

Compatibility equations constitute exchange  relations 
  among the differentials. 
They have the form of the Whitham hierarchy \cite{Whitham}: 
\begin{equation} \la{441} 
\left \{ 
\begin{array}{rcl} 
\nabla(\zeta)W^{(\xi,\bar\xi )}(z) & = & 
\nabla(\xi)W^{(\zeta,\bar\zeta )}(z), \\ 
\nabla(\zeta)W^{(\infty,\bar\infty )}(z) & = & 
\p_t W^{(\zeta,\bar\zeta )}(z), \\ 
\nabla(\zeta)W_\alpha(z) & = & \p_{t^{(\alpha )}}W^{(\zeta,\bar\zeta )}(z), 
\end{array} 
\right . 
\ee 
where $\nabla(z)=\p_{t} + 
\frac{\delta}{\delta A(z)} + 
\frac{\delta}{\delta \overline{A(z)}}$. 

Now we can check that the Whitham equations  (\ref{441}) 
are the semiclassical limit of 
the quantum deformation equations (\ref{ZS}).  Substituting 
  the semiclassical form of the wave function 
(\ref{75}) into (\ref{L2}), we see that the 
classical limit of the flows 
are primitive functions of the meromorphic differentials 
$dW^{(\xi,\bar\xi )}(z)$: 
$$H_{nm}(\xi )\to W^{(\xi,\bar\xi )}(z).$$ 
  Similarly, 
  the classical limit of the 
shift operator $\hat w$ defined in Sec.~\ref{1A} is equal to 
$W^{(\infty,\bar\infty )}(z)$, and the second equation 
of (\ref{441}) is a 
classical limit of the compatibility conditions.

The deformation equations become more illustrative 
     for algebraic domains \cite{mkwz} and especially significant 
for simply-connected algebraic domains \cite{mwz}. In these cases all 
the differentials are expressed through the Green function for the 
exterior Dirichlet boundary problem. 
Differentials of the third kind are expressed through the 
Green function on the physical sheet, 
\be\nonumber 
dW^{(\zeta,\bar\zeta )}(z)=2\p_zG(\zeta,z)dz, 
\ee 
while the holomorphic differentials are related to harmonic measures 
of the droplets: 
\be\nonumber 
dW_{\alpha}(z)=\p_z\omega_\alpha(z)dz. 
\ee 

We recall that the Green function $G(\zeta,z)$ is a 
symmetric and harmonic function 
everywhere in the exterior of $D$ except at $z=\zeta$, 
where it has a logarithmic singularity, 
$G(\zeta,z)\sim \log|\zeta-z|$. The Green function vanishes 
     if any of the arguments belongs to the boundary. 
The harmonic measure of a droplet $D_\alpha$ is 
a harmonic function in the exterior of $D$ such 
     that it is equal to $1$ on the boundary of $D_\alpha$ 
and vanishes at 
     boundaries of all other droplets. 

If there is only one physical droplet,  and no virtual droplets, 
the Green function is given 
by a conformal map $w(z)$ of 
the exterior of the droplet  to the exterior of the unit disc: 
$G(\zeta,z)=\log\big |\frac{w(\zeta)-w(z)}{1-w(\zeta) 
\overline{w(z)}}\big |$. 
In this case the deformation equations read 
\be\la{43} 
\nabla (\zeta) S(z)=-2\p_zG(\zeta,z), \quad   \quad 
\p_t S(z)=\p_z\log w(z). 
\ee 
The second equation in 
(\ref{43}) describes the evolution of the conformal 
map when the area of the domain increases, 
while all the harmonic moments 
stay constant. This equation, in different 
equivalent forms, 
has been  known in the theory of  Laplacian growth for 
a long time \cite{Kochina}. 
The first equation (\ref{43}) 
appeared in \cite{mwz,kkmwz,1}; 
for its generalization for multiply-connected 
algebraic domains, see \cite{mkz}. 
A generalization of deformation and growth equations 
for the Laplacian growth of 
multiply-connected domains was reported  in 
Ref.\cite{mkwz}. It appeares that they constitute the same 
Whitham hierarchy 
  as the semiclassical limit of the 
evolution of the normal matrix ensemble (\ref{441}).

\section{Appendix. Laplacian Growth}\label{Appendix_A} 

Laplacian growth referres to growth of a planar  domain, whose boundary 
propagates with a velocity 
proportional to the gradient of a harmonic field. The 
Hele-Shaw problem is a typical example. It describes 
the dynamics of a 2D system of two immiscible fluids (such as oil and 
water). 
The fluids are confined 
between two horizontal glass plates, 
separated by a small distance,  so that the 
problem is essentially two-dimensional.

The motion of the interface  follows from the 
Navier-Stokes equation being 
specified for a cell. It reads: velocity field of oil is proportional 
    to gradient of 
    pressure. The liquids are assumed to be incompressible, so that the 
pressure is a 
harmonic function. 

In Refrs. \cite{mwz,ABWZ02,mkwz}, we identified the Laplacian growth 
with  the Whitham hierarchy. It 
  appeares to be identical  to 
equations (\ref{441}) describing the growth and 
   deformation 
of the support of 
eigenvalues of the 
normal matrix ensemble.

\section*{Acknowledgments} 

We are indebted to 
A. Kapaev, 
V. Ka\-za\-kov, I. Kri\-che\-ver, I. Kos\-tov, 
A. Mar\-sha\-kov and M. Mi\-ne\-ev-\-Wein\-stein 
for useful discussions, interest in the subject and help. 
P.W. and R.T. were supported by the NSF MRSEC Program under 
DMR-0213745, NSF DMR-0220198 and by the Humboldt foundation. 
    A.Z. and P.W. acknowledge support 
by the LDRD project 20020006ER ``Unstable 
Fluid/Fluid Interfaces" at Los Alamos National Laboratory and M. 
Mi\-ne\-ev-\-Wein\-stein for the hospitality 
in Los Alamos. A.Z. was also supported in  part by  RFBR grant 
03-02-17373,w and by the grant for support of scientific schools 
NSh-1999.2003.2. P.W.  is grateful to   K.B. Efetov  for the hospitality in 
Ruhr-Universitaet Bochum  and to A.Cappelli for the hospitality in the 
University of Florence.

\end{document}